\documentstyle[prd,aps,floats,epsfig]{revtex}
\tighten
\begin{document}
\draft
\twocolumn[\hsize\textwidth\columnwidth\hsize\csname 
@twocolumnfalse\endcsname
\title{MODELS OF DYNAMICAL SUPERSYMMETRY BREAKING AND QUINTESSENCE.}
\author{P.~Bin\'etruy$^a$}
\address{$^a$LPTHE, Universit\'e Paris-XI, B\^atiment 211, F-91405
Orsay Cedex, France}
\maketitle
\begin{abstract}
We study several models of relevance for the dynamical breaking of
supersymmetry which could provide a scalar component with equation of
state $p=w\rho$, $-1<w<0$. Such models would provide a natural
explanation for recent data on the cosmological parameters.
\end{abstract}
\vskip2pc]

\section*{Introduction}

There are increasing indications that the energy density of matter in the 
Universe is smaller than the critical density\cite{exp}. 
If one sticks to the inflation 
prediction of $\Omega_T=1$, then the natural question is the origin of the 
extra  component providing the missing energy density. An obvious candidate is 
a cosmological constant, whose equation of state is $p=-\rho$. This faces 
particle physics with the unpleasant task of explaining why the energy of the 
vacuum should be of order $(0.003\; {\rm eV})^4$, a task possibly even harder 
than the one of explaining why the cosmological constant is zero. In 
particular, it seems to require new interactions with a typical scale 
much lower than the electroweak scale, long range interactions that would 
have remained undetected.

It has recently been proposed to consider instead a dynamical time-dependent 
and spatially inhomogeneous component, with an equation of state $p=w\rho$,
$-1<w<0$. Such a component has been named ``quintessence'' by Caldwell, Dave 
and Steinhardt \cite{CDS}.   Indeed, present cosmological data seem to prefer
\cite{TW}, in the context of cold dark matter models, a value for $w$ of 
order $-0.6$. Several candidates have been proposed for this component: 
tangled cosmic strings \cite{SP}, pseudo-Goldstone bosons \cite{FHSW}.
Of particular relevance to some issues at stake in the search for a unified 
theory of fundamental interactions is a scalar field with a scalar potential 
decreasing to zero for infinite field values \cite{CDS,ZWS}. 

It has been noted that such a behaviour appears naturally in models of
dynamical supersymmetry breaking (DSB)\cite{ZWS}. Typically, the scalar
potential of supersymmetric models has many flat directions, {\em i.e.}
directions in field space where the scalar potential vanishes. Once
supersymmetry is broken dynamically, the degeneracy corresponding to the
flat direction is lifted but generally the flat direction is restored at
infinite values of the scalar field\footnote{In some cases, the field
value may be interpreted as the inverse coupling constant associated
with the dynamics responsible for supersymmetry breaking. An infinite field
value means a vanishing gauge coupling and thus restoration of
supersymmetry.}.
We are thus precisely in the situation of a potential smoothly
decreasing to zero at infinity. This is usually considered as a drawback
of spontaneous supersymmetry breaking models from the point of view of
cosmology: in the standard approach, the potential has a stable ground
state, where the potential is fine tuned to zero (in order to account
for a vanishing cosmological constant); but the initial conditions and
the subsequent cosmological evolution
may lead to a situation where the field misses the ground state and 
evolves to infinite values.

Dynamical supersymmetry breaking  is often favoured because it
can more easily account for large mass scale hierarchies such as
$M_W/M_P$ through some powers of $\Lambda/M_P$ where $\Lambda$ is the
dynamical scale of breaking. It is thus a natural question to ask whether the
corresponding models may account for quintessence. Indeed, in this case,
there is a fundamental reason why the scalar potential vanishes at
infinity: this is related to the old result that global supersymmetry
yields a vanishing ground state energy. And there may be reasons as to
why once it dominates, the contribution of the scalar field to the
energy density is very small (again through powers of $\Lambda/M_P$).

In the following, we will discuss two models of dynamical supersymmetry
breaking which may be considered as representative of semi-realistic
models for high energy physics. One is based on gaugino condensation coupled to
the dynamics of a dilaton field, the other uses the condensation of
$N_f$ flavors in a $SU(N_c)$ gauge theory.

\section*{Models with a dilaton}

We start with a class of models, reminiscent of many superstring
models, where supersymmetry is broken through gaugino condensation 
\cite{gaugecond} along
the flat direction corresponding to the dilaton field. Indeed, in many
superstring models, the dilaton field $s$ does not appear in the
superpotential and thus corresponds to a flat direction in the scalar
potential. It couples to the gauge fields in a model-independent way: 
\begin{equation}
{\cal L} = -{1\over 4} s F^{\mu \nu} F_{\mu \nu}
\end{equation}
where $F_{\mu\nu}$ is the field strength corresponding to a generic
gauge  symmetry group $G$ and, throughout this article, $s$ is expressed
in Planck mass units. Thus the vacuum expectation value $<s>$ can 
be interpreted as the inverse of the  gauge coupling $1/g^2$ at the
string scale. Indeed, it
is directly related to the inverse of the string coupling constant
(see below). 
The interaction  corresponding to the gauge group $G$  becomes strong 
at a scale:
\begin{equation}
\Lambda = M_P e^{-1/2bg^2}=M_P e^{-s/2b_0} \label{eq:lambda}
\end{equation} 
where $b_0$ is the one-loop beta function coefficient of the gauge group $G$.
The corresponding gaugino fields are expected to condense:
\begin{equation}
<\bar \lambda \lambda> = \Lambda^3 = M_P^3 e^{-3s/2b_0}
\end{equation}
and they lead to a potential energy, quadratic in the gaugino
condensates, that scales like $e^{-3s/b_0}$. In the limit of infinite
$s$, that is of vanishing gauge coupling, the dynamics is inoperative
and one recovers the flat direction associated with the dilaton.

We have followed a very crude approach and there are, of course, many
possible refinements: one may include supergravity corrections, the
effect of other scalar fields such as moduli, as well as corrections
which may be needed to stabilize the potential for small values of $s$ 
(that is in the regime of strongly coupled string \cite{BD}). For
example, in a given model \cite{BGW}, the potential reads, in term of 
the field $\ell$ which precisely describes the string gauge coupling:
\begin{eqnarray}
V(\ell)&=&{M_P^4 \over 16 e^2 \ell}\{(1+f-l{df\over d\ell})
(1+{2\over 3}b_0 \ell)^2-{4\over
3}b_0^2\ell^2)\} \nonumber \\
& & \hskip 1cm \times \; e^{g-3(f+1)/2b_0\ell},
\end{eqnarray} 
where $f(\ell)$ and $g(\ell)$ appear as non-perturbative contributions
to the K\"ahler potential. The dilaton $s$ is related to the field
$\ell$ as $s=(1+f)/2\ell$. One recovers, in the limit of large $s$
(small string coupling $\ell$) a leading  behaviour in  $e^{-3s/b_0}$. 

Since there are obvious power law corrections to this behaviour, we will
consider a toy model of a dilaton field $s$ with a Lagrangian:
\begin{equation}
{\cal L} = -{1 \over 4 s^2} \partial^\mu s \partial_\mu s - V(s),
\end{equation}
where
\begin{equation}
V(s) = V_0(s) e^{-3s/b_0}.
\end{equation}
The non-canonical kinetic term for $s$ is caracteristic of the string
dilaton and accounts for the non-flat K\"ahler metric.

The cosmological evolution of the $s$ field is described by the
following set of equations ($\kappa = 1$):
\begin{eqnarray}
{\ddot{s} \over 2 s^2} - {\dot{s}^2 \over 2 s^3} &+& 3 H {\dot s \over 2
s^2} + {dV \over ds} = 0 \nonumber \\
H^2 &=& {1 \over 3} (\rho_B + \rho_s)
\end{eqnarray}
where $\rho_B$ is the background energy density associated with matter
($w_B=0$) or radiation ($w_B=1/3$) and $\rho_s =
\dot{s}^2/(4s^2) + V(s)$.

If we first consider that $V_0(s)$ is a constant and solve these equations 
assuming that $\rho_B$ dominates for some time, there exists a scaling
solution with the following behaviour:\footnote{ For a similar analysis,
although in a different context, see Ref. \cite{BCC}.} 
the field $s$ evolves down the
exponentially decreasing potential as $(t/t_1)^{{1- w_B \over 1+
w_B}}$ as long as $s$ remains smaller than $s_1 \equiv {2b_0 \over 3} 
{{1+ w_B \over 1- w_B}}$, reached at $t=t_1$; for larger values, 
there exists a scaling solution \cite{PR,CLW,FJ} 
where the field  evolves logarithmically as
$s=s_1 + (2b_0/3) \ln(t/t_1)$. The ratio $\rho_s/\rho_{tot}$ starts at
$3(1-w_B)^2/16$ for $t \le t_1$ and from then on slopes down to zero 
as $(b_0^2/6s^2)(1+w_B)$ for large values of
$s$. Finally, $w_s = p_s/\rho_s$ starts at a value of $1$ and
decreases monotonically towards $w_B$ as $s$ increases. There is
therefore no hope of using the dilaton for the  dynamical component
of quintessence since $w_s$ never reaches a negative value.
Power law corrections ($V_0(s) \propto s^\alpha$) do not change this
conclusion.  

This might be in some sense a welcome conclusion since the vacuum
expectation value $<s>$ provides, after renormalisation down to low
energy, the fine structure constant $1/\alpha$. A sliding dilaton would
make the fine structure constant vary with time at an unacceptable rate
\cite{alpha}. 

Similar conclusions can be reached with other types of weakly coupled
scalar particles, such as the moduli of string theories. For example, in
a model with several gaugino condensates and a modulus field $t$
describing the radius of the six dimensional compact manifold, the 
scalar potential scales for large values of $t$  as \cite{BGW}:
\begin{equation}
V= \sum_a t^{{b+b_a \over b_a}} e^{-\pi {b-b_a \over 3 b_a} t}
e^{-2<s>/b_a}
\end{equation}
where the sum runs over the different condensates (one for each group
$G_a$, with corresponding beta function coefficient $b_a$). We have
fixed the dilaton field $s$ at its ground state value. Let us note
that, although the modulus $t$ definitely 
cannot be used for quintessence (since,
as above, the corresponding $w_t$ reaches asymptotically $w_B$), a large
value of $<s>$ may contribute to giving a small contribution from $t$ to
the vacuum energy.  

\section*{A model of fermion condensates}

We now turn to a model  which yields inverse powers of fields in the
potential, a welcome situation for quintessence models \cite{ZWS}.
It is based on the gauge group $SU(N_c)$ and has $N_f\le
N_c$ flavors: quarks $Q^i,\; i=1 \cdots N_f$ in
fundamentals of $SU(N_c)$ and antiquarks  ${\tilde Q}_i,\; i=1 \cdots
N_f$  in antifundamentals of $SU(N_c)$.

Below the scale of dynamical breaking of the gauge symmetry $\Lambda$,
the effective degrees of freedom are the fermion condensate (``pion'')
fields $\Pi^i_j \equiv Q^i \tilde Q_j$. The dynamically generated
superpotential reads \cite{ADS}:
\begin{equation}
W= (N_c - N_f) {\Lambda^{{3N_c-N_f \over N_c-N_f}} \over (det \Pi)^{{1
\over N_c-N_f}}} .
\end{equation}
Usually, one allows a term linear in $\Pi$ in the superpotential in
order to stabilize this field. We will instead assume here that a
discrete symmetry ensures that no linear term is allowed by the
abelian symmetry. Let us note that this symmetry cannot be a continuous
gauge symmetry since this would yield in the scalar potential D-terms
with positive powers of $\Pi$ which would stabilize the field.

The effective Lagrangian  reads:
\begin{eqnarray}
{\cal L} &=& -{1\over 2} {\rm Tr} \; \left[ (\Pi^\dagger \Pi)^{-1/2}
\partial_\mu \Pi \partial^\mu \Pi^\dagger \right] \nonumber \\
& & + 2 {\rm Tr} \; \left[  (\Pi^\dagger \Pi)^{-1/2}{(\Lambda
\Lambda^\dagger)^{{3N_c-Nf \over N_c-N_f}} \over ({\rm Det} \;
\Pi^\dagger \Pi)^{{1 \over N_c-N_f}}} \right]
\end{eqnarray}
where the potential originates from the F-term for the field $\Pi$.
For simplicity, we will take   $\Pi^i_j$ to be diagonal and write
$\Pi^i_j \equiv \Phi^2 \delta^i_j$ with $\Phi$ real. One obtains:

\begin{equation}
{1\over 4N_f}{\cal L} = -{1 \over 2} \partial^\mu \Phi \partial_\mu \Phi + V(\Phi)
\end{equation}
where
\begin{equation}
V(\Phi) = \lambda {\mu^{4+\alpha} \over \Phi^\alpha},
\end{equation}
with $\mu = (\Lambda \Lambda^\dagger)^{1/2}$ and
\begin{equation}
 \alpha = 2 {N_c + N_f \over N_c - N_f}.
\end{equation}
The corresponding potential has been studied in Ref.\cite{PR} in the
case where $\rho_B$ dominates over the energy density $\rho_\Phi$ of the
$\Phi$ field. One obtains
\begin{equation}
{\rho_\Phi \over \rho_B} = \left( {a \over a_Q}
\right)^{6(1+w_B)/(2+\alpha)}.
\end{equation}
Hence $\rho_\Phi$ decreases less rapidly than $\rho_B$ until it
dominates it for values of the cosmic scale factor larger than
$a_Q$. Throughout this period (which must obviously include
nucleosynthesis), one has:
\begin{eqnarray}
\rho_\Phi &=& {2(2+\alpha) \over 4+ \alpha(1-w_B)} \left({3(1+w_B) \over
\alpha(2+\alpha) }\right)^{{\alpha\over 2}} \lambda {\mu^{4+\alpha} \over
M_P^\alpha} \left({a\over a_Q}\right)^{{6(1+w_B)\over 2+\alpha}} \nonumber
\\
\Phi &=& M_P \sqrt{{\alpha(2+\alpha) \over 3(1+w_B)}} \left({a \over
a_Q}\right)^{{3(1+w_B)\over 2+\alpha}}. 
\end{eqnarray}
The equation of state for the $\Phi$ field has \cite{ZWS}:
\begin{equation}
w_\Phi = -1 + {\alpha (1+ w_B)  \over 2+\alpha}. \label{eq:wphi}
\end{equation}
Thus, in a matter-dominated universe ($w_B=0$), $w_\Phi =
-1/2+2N_f/N_c$ which is  between $-1/2$ and $0$ for $N_f \le
N_c$. This provides a candidate for the dynamics of quintessence.

Once $\Phi/M_P$ has reached the value $\sqrt{{\alpha(2+\alpha)\over
3(1+w_B)}}$, we enter a different regime where $\rho_\Phi$
dominates the energy density. The field $\Phi$ slows down and one may
solve for ir neglecting the terms $\ddot{\Phi}$ in its equation of
motion and $\dot{\Phi}^2/2$ in $\rho_\Phi$. One obtains:
\begin{equation}
\Phi = \Phi_0 \left[ 1 + {1\over 2 \sqrt{3}} \alpha (4+\alpha) V(\Phi_0)
(t-t_0)\right]^{{2\over 4+\alpha}},
\end{equation}
where $\Phi_0$ is the present value for $\Phi$, and one obtains
\begin{equation}
w_\Phi \sim -1 + {\alpha^2 \over 3 \Phi^2}.
\end{equation}
If $\rho_\Phi$ at $a_Q$ is already close to the present value (this
occurs typically for $\mu \sim 10^{-12+30\alpha/(4+\alpha)}$ GeV), 
this second period is short ($a_Q \sim a_0$)
and $w_\phi$ will be given approximately by (\ref{eq:wphi}). For
simplicity, we will suppose from now on that this is so. In this case,
the value of $w_\Phi$ might prove to be too small to account for the
data \cite{SCP}.

However, larger values for $w_\Phi$ may be obtained by complicating
slightly the model and introducing other fields. As an example, we will
assume the presence of a dilaton field, much in the spirit of the
models of the previous section (although the dilaton is this time 
not sliding but
fixed at its ground state value). The dynamical scale $\Lambda$ is
expressed in terms of the dilaton through (\ref{eq:lambda})
with $b_0 = (3 N_c-N_f)/(16 \pi^2)$. This induces a new term in the
scalar potential:
\begin{equation}
\delta V= 4 s^2 |F_s|^2,
\end{equation}
with
\begin{equation}
F_s = {dW\over ds} = -8 \pi^2 {\Lambda^{{3N_c-N_f \over N_c-N_f}}\over
({\rm Det}\; \Pi)^{{1 \over N_c-N_f}}}.
\end{equation}
that is an extra term of the form $\mu^{4+\beta}/\Phi^\beta$ with
\begin{equation}
\beta = {4N_f \over N_c-N_f}.
\end{equation}
Since $\beta<\alpha$, this term dominates for large values of the
condensate $\Phi$ and, for $w_B=0$,
\begin{equation}
w_\Phi  = -1 +{2N_f \over N_c+N_f},
\end{equation}
which precisely lies between $-1$ and $0$: taking for example $N_c=5$
and $N_f=1$ yields $w_\Phi = -2/3$.

There could be other contributions to the $F$-term auxiliary field for
$S$, say $F_0$ (which will contribute to supersymmetry breaking). If so,
the leading term in $\delta V$ for large $\Phi$ is $F_s^\dagger F_0 +
F_s F_0^\dagger$ and $\beta=2N_c/(N_c-N_f)$, in which case $w_\Phi =
-1+ N_f/N_c$. This time, one may even obtain $w_\Phi= -2/3$ with $N_c=3$
($N_f=1$).

Strictly speaking, the leading term is $|F_0|^2$ and thus of the
cosmological constant type. But this is an artifact of global
supersymmetry and it is well-known that, by going to supergravity, we
may cancel this cosmological constant term, while keeping a
non-vanishing contribution $F_0$ to the $F$-term of the $S$ field.
Such a study goes beyond the framework of this paper. This stresses
however an important fact: even if we deal here with a dynamical
component ($\Phi$) which may account for a cosmological constant type
behaviour of the cosmological parameters, it is important that the
$\Phi$ energy density eventually dominates over all other forms and 
thus that these other components do not produce a significant 
cosmological constant of their own.
Thus, the cosmological constant remains a problem for all
other components.

Likewise, the amount of supersymmetry breaking due to the fact that
$\Phi$ has not reached an infinite value (and thus its $F$-term is not
vanishing) is not sufficient to account for the amount of
supersymmetry-breaking observed in nature. There must be other sources
({\em e.g.} $F_0$ in our example) which may produce unwanted amounts of
cosmological constant if care is not taken.

In other words, there is still a ``cosmological constant problem''  in
the models studied here (that is to say, from the point of view of the 
quantum theory) but the interest of such models lies in the fact that they can
successfully account for the recent cosmological data on supernovae of
type Ia, if confirmed. 

\vskip .8cm
{\bf Acknowledgments}
\vskip .5cm
I wish to thank Alex Vilenkin for raising my interest in the issue of
quintessence, the Berkeley Lab Theory group  for hospitality while
part of this work was done, and Reynald Pain for valuable discussions.

\end{document}